\documentclass[aps,amsfonts]{revtex4}

\usepackage{epsfig,psfrag}
\usepackage{latexsym}
\usepackage{graphicx}
\def\beq{\begin{eqnarray}}
\def\eeq{\end{eqnarray}}
\def\bea{\begin{eqnarray}}
\def\eea{\end{eqnarray}}

\usepackage{color}

\begin{document}

\title{Abelian Zero  Modes in Odd Dimensions}
\author{Gerald V. Dunne}\email{dunne@phys.uconn.edu}
\affiliation{Department of Physics, University of Connecticut, Storrs, CT 06269}
\author{Hyunsoo Min}\email{hsmin@dirac.uos.ac.kr}
\affiliation{Department of Physics,  University of Seoul, Seoul 130-743, Korea}

%\date{\today}

\begin{abstract}
We show that the Loss-Yau zero modes of the 3d abelian Dirac operator may be interpreted in a simple manner in terms of a stereographic projection from a 4d Dirac operator with a constant  field strength of definite helicity. This is an alternative to the conventional viewpoint involving Hopf maps from $S^3$ to $S^2$. Furthermore, our construction generalizes in a straightforward way to any odd dimension. The number of zero modes is related to the Chern-Simons number in a nonlinear manner.
\end{abstract}

\maketitle

%\subsection{Introduction}

The behavior of quantized charged fermions in ultra-strong magnetic fields has applications in atomic physics \cite{loss}, particle and condensed matter physics \cite{gusynin}, and astrophysics \cite{field}. Key properties determining stability are the existence of zero modes for the 3 dimensional Dirac operator, and the associated magnetic helicity. In studying the stability of atoms in magnetic fields, Loss and Yau \cite{loss} found the surprising result that the abelian Dirac operator in 3 dimensions, ${\mathcal D}_3\equiv i\gamma_\mu\left(\partial/\partial x^\mu-i A_\mu \right)$, can have exact zero modes 
\begin{eqnarray}
 i\gamma_\mu\left(\frac{\partial}{\partial x^\mu}-i A_\mu\right)\,\psi^{(0)}=0
 \label{zm}
\end{eqnarray}
for smooth, localized magnetic fields $\vec{B}=\vec{\nabla}\times \vec{A}$. 
In even dimensions there is a well-known relation between zero modes and the topology of gauge fields \cite{current}, but in odd dimensions, where the relevant index theorem is due to Callias \cite{callias}, the situation is somewhat different, as the index is determined by the topology of the coupling to a Higgs field. In this Brief Report we present a simple new interpretation of the Loss-Yau zero-mode-supporting abelian gauge fields, and show that this construction generalizes to all odd dimensions.

\begin{figure}
\includegraphics[scale=0.8]{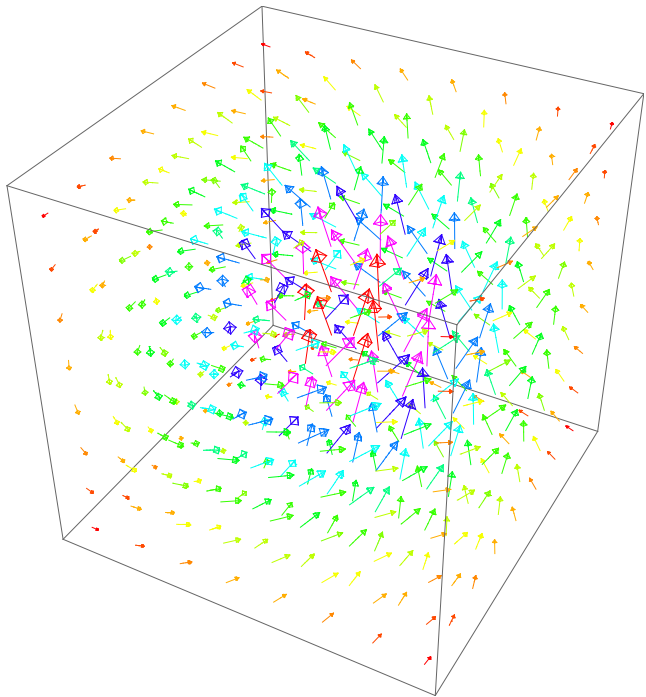}
\caption{Plot of the magnetic field vector $\vec{B}_{LY}$ in (\ref{ly-gauge}). Note that the magnitude is highly localized around the origin, while the direction winds in a non-trivial manner.}
\label{fig1}
\end{figure}

Loss and Yau's simplest example \cite{loss} is the gauge field
\begin{eqnarray}
\vec{A}_{LY}=\frac{3}{1+\vec{x}^2}\,\hat{N}\quad , \quad \hat{N}\equiv\frac{1}{1+\vec{x}^2} 
\begin{pmatrix}
%{2y_1y_3-2y_2\cr
%2y_2y_3+2y_1\cr
%1-y_1^2-y_2^2+y_3^2}
{2x_1 x_3-2x_2\cr
2x_2 x_3+2x_1\cr
1-x_1^2-x_2^2+x_3^2}
\end{pmatrix}
\quad , \quad \vec{B}_{LY}=\frac{12}{\left(1+\vec{x}^2\right)^2}\,\hat{N} \quad ,
\label{ly-gauge}
\end{eqnarray}
for which the zero mode is
\begin{eqnarray}
\psi_{LY}^{(0)}=\frac{4}{\left(1+\vec{x}^2\right)^{3/2}}\left(1+i\gamma_\mu x_\mu\right) 
\begin{pmatrix}
{1\cr
0}
\end{pmatrix} \quad .
\label{ly-zero}
\end{eqnarray}
For this field, the Chern-Simons number, or magnetic helicity, is
\begin{eqnarray}
{\mathcal N}^{CS}_{LY}\equiv \frac{1}{16\pi^2}\int d^3 x\, \vec{A}\cdot\vec{B}=\frac{9}{16} \quad .
\label{ly-cs}
\end{eqnarray}
The associated magnetic field is plotted in Figure \ref{fig1}, showing the localized but non-trivial structure of the field. 
%[side comment: Note that in the Loss-Yau gauge, $\vec{A}_{LY}$ is locally directed along the lines of $\vec{B}_{LY}$; in other gauges this is not the case [an interesting other gauge is that with $A_{3}=0$, where one finds that the vector potential interpolates between a source and a sink as a function of $x_3$].]
This basic example may be extended \cite{loss,adam} to fields with multiple zero modes:
\begin{eqnarray}
\vec{A}_{LY}=\frac{(2k+3)}{1+\vec{x}^2}\,\hat{N} \quad ; \quad
\vec{B}_{LY}=\frac{4(2k+3)}{\left(1+\vec{x}^2\right)^2}\,\hat{N} \quad ; \quad
%\psi_{LY}^{(0)}&=& \left(1+i\gamma_\mu x_\mu\right) (...) \quad, \quad m= 1....\\
{\mathcal N}^{CS}_{LY}&=& \frac{(2k+3)^2}{16} \quad .
 \label{ly-deg}
\end{eqnarray}
Here $k\geq 0$ is an integer, and this field has $(k+1)$ zero modes that can be expressed in terms of 3d spherical harmonics \cite{loss,adam}.

These fields have since been discussed in terms of Hopf maps \cite{adam,erdos}, which are maps $\chi: S^3\to S^2$ such that the magnetic field is
\begin{eqnarray}
\vec{B}_{H}=\frac{2}{i}\frac{\vec{\nabla}\chi\times \vec{\nabla} \bar{\chi}}{\left(1+\chi\bar{\chi}\right)^2} \quad .
\label{hopf-b}
\end{eqnarray}
Such a Hopf map can also be viewed as a map $\chi:{\bf R}^3\to {\bf R}^2$, and the simplest example
\begin{eqnarray}
\chi=\frac{(x_1+ix_2)}{2x_3-i(1-\vec{x}^2)}\quad ; \quad \vec{B}_H=\frac{16}{\left(1+\vec{x}^2\right)^2}\,\hat{N}=\frac{4}{3}\vec{B}_{LY}\quad ; \quad {\mathcal N}_{H}^{CS}=1
\label{hopf-1}
\end{eqnarray}
gives a magnetic field proportional to the Loss-Yau field in (\ref{ly-gauge}).
Geometrically, $\vec{B}_H$ is tangent to the closed curves in ${\bf R}^3$ given by $\chi={\rm constant}$. Erd\"os and Solovej \cite{erdos}
 gave an elegant interpretation of these zero-mode-supporting gauge fields  in terms of pull-backs (to ${\bf R}^3$) of 2 dimensional magnetic fields, and the 3 dimensional zero modes were related to the Aharonov-Casher zero modes in 2 dimensions \cite{ac}. 
% Higher degeneracies may be generated for more complicated Hopf maps, and by combining the Hopf map with rational maps. 
Further results have been found in \cite{adam,elton}, and 
these gauge fields have also been understood in terms of projections of non-abelian fields \cite{jackiw}. 

However, a number of questions remain. The fundamental mis-match of the coefficient [the factor 4/3 in (\ref{hopf-1})] does not have an elegant interpretation in the Hopf map language. In one picture, one introduces an additional "background" field with a correcting coefficient \cite{adam}; and in another picture \cite{erdos}, one includes a magnetic monopole of  a particular strength at the centre of the $S^2$ to adjust the strength of the area form. 

In this short note we present another characterization of these 3d abelian zero-mode-supporting gauge fields, in terms of four dimensional gauge fields of fixed helicity. This construction is extremely simple, and furthermore it generalizes naturally to other odd dimensions.

%\subsection{Gauge field via stereographic projection}

Our basic example [the analogue of (\ref{ly-gauge})] is expressed for arbitrary odd dimension by a stereographic projection from ${\bf R}^{2n}\supset S^{2n-1} \to {\bf R}^{2n-1}$. We define coordinates $x_\mu$ ($\mu=1, 2, \dots , 2n-1$) on ${\bf R}^{2n-1}$, and coordinates $y_a$ ($a=1,\dots , 2n$) on ${\bf R}^{2n}$. Consider a $2n$-dimensional gauge field corresponding to a constant field strength, and such that the field has fixed helicity :
\begin{eqnarray}
%{\mathcal A}_a&=&-\frac{\mathcal F}{2}\left(-y_2, y_1, -y_4, y_3, \dots , -y_{2n}, y_{2n-1}\right)
{\mathcal A}_a&=&-\frac{\mathcal F}{2}\left(y_2, -y_1, y_4, -y_3, \dots , y_{2n-2},-y_{2n-3},-y_{2n}, y_{2n-1}\right)
\nonumber\\
&\equiv& -\frac{\mathcal F}{2} J_{ab}\, y_b
\label{sd-gauge}
\end{eqnarray}
where the antisymmetric matrix $J={\rm diag}(i\sigma_2, \dots, i\sigma_2, -i\sigma_2)$. [The sign-flip in the last diagonal entry is a parity convention chosen to agree with the choice of Loss-Yau.]
This gauge field is in Fock-Schwinger gauge: $y_a{\mathcal A}_a=0$. 
Now restrict to $S^{2n-1}$ by imposing the condition $y_a^2=1$, and stereographically project from $S^{2n-1}$ to ${\bf R}^{2n-1}$ via:
\begin{eqnarray}
%y_\mu=\frac{2x_\mu}{1+r^2} \quad , \quad y_{2n}=\frac{1-r^2}{1+r^2}\quad, \quad r^2\equiv x_\mu^2
y_\mu=\frac{2x_\mu}{1+\vec{x}^2} \quad , \quad y_{2n}=\frac{1-\vec{x}^2}{1+\vec{x}^2} \quad .
%, \quad r^2\equiv x_\mu^2
\label{stereo}
\end{eqnarray}
The projected $(2n-1)$-dimensional gauge field $A_\mu$ is \footnote{Such gauge field projections have been studied extensively for projections to ${\bf R}^4$ \cite{adler,jackiwrebbi,bogomolny}; here we consider analogous projections to odd dimensional spaces.}
\begin{eqnarray}
A_\mu=\frac{\partial y_a}{\partial x_\mu}{\mathcal A}_a
\label{projected-gauge}
\end{eqnarray}
One finds by a simple computation 
\begin{eqnarray}
%A_i&=& 2 {\mathcal F} \left(\frac{J_{ij} x_j+x_i x_{2n-1}}{(1+x^2)^2}\right)\quad ,\quad i=1, 2, \dots 2n-2 \\
%A_{2n-1}&=&2 {\mathcal F} \left(\frac{1-x^2+2x_{2n-1}^2}{(1+x^2)^2}\right)
%A_i&=& 2 {\mathcal F} \left(-\frac{J_{ij} x_j+x_i x_{2n-1}}{(1+r^2)^2}\right)\quad ,\quad i=1, 2, \dots 2n-2 \\
%A_{2n-1}&=& {\mathcal F} \left(\frac{1-r^2+2x_{2n-1}^2}{(1+r^2)^2}\right)
A_i&=& 2 {\mathcal F} \left(\frac{-J_{ij} x_j+x_i x_{2n-1}}{(1+\vec{x}^2)^2}\right)\quad ,\quad i=1, 2, \dots 2n-2 \nonumber\\
A_{2n-1}&=& {\mathcal F} \left(\frac{1-\vec{x}^2+2x_{2n-1}^2}{(1+\vec{x}^2)^2}\right)
\label{gauge-p}
\end{eqnarray}
When $n=2$ (i.e.,  a 3 dimensional gauge field $A_\mu$) this reproduces precisely the form of the original Loss-Yau gauge field \cite{loss} in (\ref{ly-gauge}), although the coefficient ${\mathcal F}$ is not yet determined.

%\subsection{Gauge field from the zero mode}

The coefficient ${\mathcal F}$ is fixed by an explicit construction of the zero mode, directly from the zero mode equation (\ref{zm}). Straightforward Dirac algebra manipulations \cite{loss,jarvis} show that the gauge field in (\ref{zm}) can be expressed in terms of the zero mode $\psi_{(0)}$
% \footnote{This is the abelian version of the familiar instanton analysis \cite{jackiwrebbi}.}
\begin{eqnarray}
A_\mu=\frac{\partial_\nu\left(\psi_{(0)}^\dagger \sigma_{\mu\nu}\psi_{(0)}\right)}{\psi_{(0)}^\dagger \psi_{(0)}}+i\frac{\left(\partial_\mu\psi_{(0)}^\dagger \psi_{(0)}-\psi_{(0)}^\dagger \partial_\mu \psi_{(0)}\right)}{2\psi_{(0)}^\dagger \psi_{(0)}}
\label{gauge}
\end{eqnarray}
where $\sigma_{\mu\nu}=\frac{1}{4i}[\gamma_\mu, \gamma_\nu]$, with the constraint $\partial_\mu\left(\psi_{(0)}^\dagger \gamma_\mu \psi_{(0)}\right)=0$.

For our gauge field (\ref{gauge-p}) we postulate the (un-normalized) $2^{n-1}$ component  Dirac zero mode spinor  
\begin{eqnarray}
\psi_{(0)}=\frac{({\bf 1}+i\gamma_\mu x_\mu)}{(1+\vec{x}^2)^{n-1/2}} \, (1, \dots, 1 | 0, \dots, 0)^T
%\begin{pmatrix}
%{1\cr \vdots\cr 1\cr --\cr 0\cr\vdots\cr 0 
%}
%\end{pmatrix}
\label{zero}
\end{eqnarray}
Then the resulting gauge field constructed from (\ref{gauge}) in $(2n-1)$ dimensions is
\begin{eqnarray}
%A_i&=& 2 (2n-1) \left(\frac{J_{ij} x_j+x_i x_{2n-1}}{(1+x^2)^2}\right)\quad ,\quad i=1, 2, \dots 2n-2 \\
%A_{2n-1}&=&2(2n-1)\left(\frac{1-x^2+2x_{2n-1}^2}{(1+x^2)^2}\right)
A_i&=& 2 (2n-1) \left(\frac{-J_{ij} x_j+x_i x_{2n-1}}{(1+\vec{x}^2)^2}\right)\quad ,\quad i=1, 2, \dots 2n-2 \\
A_{2n-1}&=& (2n-1)\left(\frac{1-\vec{x}^2+2x_{2n-1}^2}{(1+\vec{x}^2)^2}\right)
\label{gauge-z}
\end{eqnarray}
which is precisely the same as (\ref{gauge-p}), but now the overall coefficient has been fixed to be ${\mathcal F}=2n-1$. When $n=2$ (corresponding to 3 dimensions) this reproduces  the original Loss-Yau gauge field  in (\ref{ly-gauge}). Note that the zero mode (\ref{zero}) is normalizable in all odd dimensions $d=2n-1\geq 3$.

%\subsection{Chern-Simons terms}

%The abelian Chern-Simons number in $d=2n-1$ dimensions is
%\begin{eqnarray}
%{\mathcal N}^{CS}=\frac{1}{n! (2\pi)^{n}}\int d^{2n-1}x\,  \epsilon_{\mu_1\mu_2\dots \mu_{2n-1}} A_{\mu_1} F_{\mu_2\mu_3}\dots F_{2n-2,2n-1}
%\end{eqnarray}
%For the gauge field in (\ref{gauge-z}) this gives
%\begin{eqnarray}
%{\mathcal N}^{CS}= ... (d)^{(d+1)/2}  ???
%\end{eqnarray}
%${\mathcal N}^{CS}$ is nonlinear in $n$ simply because each factor of $A$ and $F$ contributes a factor of $(2n-1)$. When $n=2$ we recover the Chern-Simons number (\ref{ly-cs}) of the original Loss-Yau field (\ref{ly-gauge}).

%\subsection{Degeneracy of zero  modes}

The degeneracy of the abelian zero modes can be deduced by  group theoretic arguments for spinors in arbitrary dimensions. Define  $2^{n-1}\times 2^{n-1}$ Dirac matrices $\gamma_\mu$ ($\mu=1, \dots, (2n-1)$) for ${\bf R}^{2n-1}$, and $2^{n}\times 2^{n}$ Dirac matrices $\Gamma_a$ ($a=1, \dots, 2n$) for ${\bf R}^{2n}$. These can be related as
\begin{eqnarray}
\Gamma_\mu =\begin{pmatrix}
{0 & i \gamma_\mu\cr
-i \gamma_\mu &0} 
\end{pmatrix}
\quad;\quad \Gamma_{2n} =\begin{pmatrix}
{0 & {\bf 1}\cr
{\bf 1} &0} 
\end{pmatrix}
\quad;\quad \Gamma_{2n+1} =\begin{pmatrix}
{{\bf 1} & 0\cr
0 & -{\bf 1}} 
\end{pmatrix}
\end{eqnarray}
Then the spin matrices in ${\bf R}^{2n}$ may be block-decomposed as
\begin{eqnarray}
\Sigma_{ab}\equiv \frac{1}{4i}[\Gamma_a, \Gamma_b] =\begin{pmatrix}
{\Sigma_{ab}^{+} &0\cr
0& \Sigma_{ab}^{-}}
\end{pmatrix} 
\end{eqnarray}
%\begin{eqnarray}
%\Sigma_{ab}&\equiv& \frac{1}{4i}[\Gamma_a, \Gamma_b] =\begin{pmatrix}
%{\Sigma_{ab}^{+} &0\cr
%0& \Sigma_{ab}^{-}}
%\end{pmatrix} \nonumber\\
%\Sigma_{\mu\nu}^{\pm}&=&\sigma_{\mu\nu}\equiv \frac{1}{4i}[\gamma_\mu, \gamma_\nu] \nonumber\\
%\Sigma_{\mu, 2n}^{\pm}&=&\pm \frac{1}{2}\,\gamma_\mu
%\end{eqnarray}
where $\Sigma_{\mu\nu}^{\pm}=\sigma_{\mu\nu}\equiv \frac{1}{4i}[\gamma_\mu, \gamma_\nu]$, and $
\Sigma_{\mu, 2n}^{\pm}=\pm \frac{1}{2}\,\gamma_\mu$.
We also define the $2n$ dimensional angular momentum generators 
\begin{eqnarray}
L_{ab}\equiv -i\left(y_a \frac{\partial}{\partial y_b}-y_b \frac{\partial}{\partial y_a}\right)
\label{orbital-2n}
\end{eqnarray}
Then a canonical result of stereographic projection of the {\it free} Dirac equation from $S^{2n-1}$ (defined by $y_a y_a=1$) to ${\bf R}^{2n-1}$ is that
\begin{eqnarray}
\left(\frac{1+\vec{x}^2}{2}\right)^{2n-1} i\gamma_\mu \frac{\partial}{\partial x_\mu}=\left(\frac{1+\vec{x}^2}{2}\right)^{n-3/2}V^\dagger \left[\Sigma_{ab}^+ L_{ab}+\left(n-\frac{1}{2}\right){\bf 1} \right]V \left(\frac{1+\vec{x}^2}{2}\right)^{n-3/2}
\label{free-dirac}
\end{eqnarray}
where $V\equiv \frac{1}{\sqrt{2}}\left({\bf 1}+i \gamma_\mu x_\mu\right)$.

Now we observe that for the $2n$ dimensional gauge field ${\mathcal A}_a$ defined in (\ref{projected-gauge}) and the $(2n-1)$-dimensional gauge field $A_\mu$ defined in (\ref{gauge-z}), this projection property of the free Dirac equation is maintained once the gauge field is included:
%\begin{eqnarray}
%\left(\frac{1+\vec{x}^2}{2}\right)^{2n-1} i\gamma_\mu \left(\frac{\partial}{\partial x_\mu}-iA_\mu\right)&=&\left(\frac{1+\vec{x}^2}{2}\right)^{n-3/2}V^\dagger \left[\Sigma_{ab}^+ {\mathcal L}_{ab}+
%\left(n-\frac{1}{2}\right){\bf 1} \right]V \left(\frac{1+\vec{x}^2}{2}\right)^{n-3/2} \nonumber\\ \\
%{\mathcal L}_{ab}&\equiv&L_{ab}+\left(y^a{\mathcal A}_b-y^b{\mathcal A}_a\right)
%\label{full-dirac}
%\end{eqnarray}
\begin{equation}
\left(\frac{1+\vec{x}^2}{2}\right)^{2n-1} i\gamma_\mu \left(\frac{\partial}{\partial x_\mu}-iA_\mu\right)=\left(\frac{1+\vec{x}^2}{2}\right)^{n-3/2}V^\dagger \left[\Sigma_{ab}^+ {\mathcal L}_{ab}+
\left(n-\frac{1}{2}\right){\bf 1} \right]V \left(\frac{1+\vec{x}^2}{2}\right)^{n-3/2} 
%{\mathcal L}_{ab}&\equiv&L_{ab}+\left(y^a{\mathcal A}_b-y^b{\mathcal A}_a\right)
\label{full-dirac}
\end{equation}
where ${\mathcal L}_{ab}\equiv L_{ab}+\left(y^a{\mathcal A}_b-y^b{\mathcal A}_a\right)$. Thus, the zero-mode equation on ${\bf R}^{2n-1}$ can be lifted to a zero-mode equation on $S^{2n-1}$, where the solutions can be written in terms of the spinor spherical harmonics in ${\bf R}^{2n}$.

To illustrate this explicitly we consider the $n=2$ case. The $4$-dimensional gauge field may be written as ${\mathcal A}_a=-{\mathcal F}/2 \bar{\eta}_{ab}^3 y_b$, where $ \bar{\eta}_{ab}^3$ is the 3rd isospin component of the standard $4$-dim. 't Hooft tensor \cite{thooft,jackiwrebbi}. 
%Then
%\begin{eqnarray}
%\left(\frac{1+\gamma_5}{2}\right)\Sigma_{ab}^{+}L_{ab}=4\vec{S}\cdot\vec{L}
%\end{eqnarray}
%where $\vec
The $4$-component spinor zero mode $\psi_{(0)}$ in ${\bf R}^4$ may be written in terms of a $2$-component spinor $u$ of definite (we choose positive) helicity:
\begin{eqnarray}
\psi_{(0)}=\begin{pmatrix}
{u\cr
0}
\end{pmatrix}
\end{eqnarray}
Then using (\ref{full-dirac}) the zero mode equation (\ref{zm}) becomes an algebraic equation
\begin{eqnarray}
%\left[\Sigma_{ab}^{+}{\mathcal L}_{ab}+3/2\right] \phi=\left(4\vec{S}\cdot\vec{L}+3/2 -\frac{{\mathcal F}}{2}\sigma_3\right)\phi=0
\left[\Sigma_{ab}^{+}{\mathcal L}_{ab}+3/2\right] u =\left(4\vec{S}\cdot\vec{L}+3/2 -\frac{{\mathcal F}}{2}\sigma_3\right)u=0
\end{eqnarray}
where $u=((1+\vec{x}^2)/2)^{1/2} V\phi$, and where $\vec{S}$ and $\vec{L}$ are angular momentum operators, of spin $1/2$ and $l$ (= half integer), respectively. We define the total angular momentum $\vec{J}=\vec{S}+\vec{L}$, and use the spinor spherical harmonics \cite{pais}
\begin{eqnarray}
u^{(\pm)}=\frac{1}{\sqrt{2l+1}}
\begin{pmatrix}
{\pm \sqrt{l+1/2\pm M} \,Y^{l}_{m, M-1/2}\cr
\sqrt{l+1/2\mp M}\, Y^{l}_{m,M+1/2}}
\end{pmatrix}
\end{eqnarray}
%with $j=l\pm 1/2$, $-(l+1/2)\leq M\leq (l+1/2)$, and $-l\leq m\leq l$. Then 
with $j=l\pm 1/2$, $-j\leq M\leq j$, and $-l\leq m\leq l$. Then 
\begin{eqnarray}
4\vec{S}\cdot\vec{L} \, u^{(\pm)}=
\left\{
\begin{matrix}
{2l\cr
-2l-2}
\end{matrix}
\right\}
u^{(\pm)} .
\end{eqnarray}
Thus  a zero mode can only occur when
$j=M=l+1/2$ and
\begin{eqnarray}
2l+3/2-\frac{{\mathcal F}}{2}=0 \quad \rightarrow \quad {\mathcal F}=4l+3
\end{eqnarray}
for arbitrary value of $-l\leq m\leq l$.  In this case $u^{(+)}=\begin{pmatrix}{1\cr 0}\end{pmatrix} Y^{l}_{m,l}$.
Recalling that $l=k/2$ is a half-integer, we arrive at the general Loss-Yau case (\ref{ly-deg}). Furthermore, the degeneracy is simply given by $2l+1=k+1$, as found by Loss-Yau in \cite{loss}.
This construction makes it clear why the zero mode degeneracy factor is linear in the integer $k$, while the Chern-Simons number is quadratic in $k$. An analogous construction is clearly  possible in higher dimensions using generalized spinor spherical harmonics, but we do not present the details here.

%\subsection{Conclusion}

To conclude, we have given a simple new interpretation of the Loss-Yau abelian zero-mode-supporting gauge fields in three dimensions, and have extended the construction to obtain new zero-mode-supporting abelian gauge fields in other odd dimensions. An interesting outstanding problem is the possibility of including an interaction with a scalar field, which might shed light on the possible relation of these fields to the Callias index theorem \cite{callias}.

\bigskip

GD thanks the US DOE for support through grant DE-FG02-92ER40716, and HM thanks the UConn Research Foundation and the USeoul Research Foundation for grants.

%\bibliographystyle{plain}
%\bibliography{addarchive,addbook,archive,addunpub}

\begin{thebibliography}{12345}

\bibitem{loss}
M.~Loss and H-T.~Yau,
"Stability of Coulomb Systems with Magnetic Fields'',
Comm.\ Math.\ Phys.\ {\bf 104}, 283 (1986).

%\bibitem{gusynin}
%  V.~P.~Gusynin, V.~A.~Miransky and I.~A.~Shovkovy,
%  ``Catalysis of dynamical flavor symmetry breaking by a magnetic field in
%  (2+1)-dimensions,''
%  Phys.\ Rev.\ Lett.\  {\bf 73}, 3499 (1994)
%  [Erratum-ibid.\  {\bf 76}, 1005 (1996)]
%  [arXiv:hep-ph/9405262].
%  %%CITATION = PRLTA,73,3499;%%

\bibitem{gusynin}
  V.~P.~Gusynin, V.~A.~Miransky and I.~A.~Shovkovy,
  ``Dynamical chiral symmetry breaking by a magnetic field in QED,''
  Phys.\ Rev.\  D {\bf 52}, 4747 (1995)
  [arXiv:hep-ph/9501304].
  %%CITATION = PHRVA,D52,4747;%%

\bibitem{field}
  G.~B.~Field and S.~M.~Carroll,
  ``Cosmological magnetic fields from primordial helicity,''
  Phys.\ Rev.\  D {\bf 62}, 103008 (2000)
  [arXiv:astro-ph/9811206].
  %%CITATION = PHRVA,D62,103008;%%

\bibitem{current}
S.~B.Trieman, R.~Jackiw, B.~Zumino and E.~Witten (Eds.), {\it Current Algebras and Anomalies} (Princeton Univ. Press, 1985).

\bibitem{callias}
  C.~Callias,
  ``Index Theorems On Open Spaces,''
  Commun.\ Math.\ Phys.\  {\bf 62}, 213 (1978).
  %%CITATION = CMPHA,62,213;%%
 
%\cite{Adam:1999mq}
\bibitem{adam}
  C.~Adam, B.~Muratori and C.~Nash,
  ``Zero modes of the Dirac operator and the Seiberg-Witten equations in  three
  dimensions,''
  Phys.\ Rev.\  D {\bf 60}, 125001 (1999)
  [arXiv:hep-th/9903040];
  %%CITATION = PHRVA,D60,125001;%%
 %\cite{Adam:1999ih}
%      \bibitem{Adam:1999ih}
%        C.~Adam, B.~Muratori and C.~Nash,
        ``Degeneracy of zero modes of the Dirac operator in three dimensions,''
        Phys.\ Lett.\  B {\bf 485}, 314 (2000)
        [arXiv:hep-th/9910139];
        %%CITATION = PHLTA,B485,314;%%
%\bibitem{Adam:2000vf}
%  C.~Adam, B.~Muratori and C.~Nash,
  ``Multiple zero modes of the Dirac operator in three dimensions,''
  Phys.\ Rev.\  D {\bf 62}, 085026 (2000)
  [arXiv:hep-th/0001164].
  %%CITATION = PHRVA,D62,085026;%%
  
  \bibitem{erdos}
  L.~Erd\"os and J.~P.~Solovej,
  ``The kernel of Dirac operators on $S^3$ and $R^3$'',
  Rev.\ Math.\ Phys.\ {\bf 13}, 1247 (2001).
  [arXiv:math-ph/0001036].
  
  \bibitem{ac}
  Y.~Aharonov and A.~Casher,
  ``The Ground State Of A Spin 1/2 Charged Particle In A Two-Dimensional
  Magnetic Field,''
  Phys.\ Rev.\ A {\bf 19}, 2461 (1979).
    %%CITATION = PHRVA,A19,2461;%%
  %%CITATION = TAUP-680-78;%%
  
   \bibitem{elton}
  D.~M.~Elton, ``New examples of  zero modes'', J.\ Phys.\ A {bf 33}, 7297 (2000),
  ``The Local Structure of Zero Mode Producing Magnetic Potentials'',
  Comm.\ Math.\ Phys.\ {\bf 229}, 121 (2002).
  
  %\cite{Jackiw:1999bd}
\bibitem{jackiw}
  R.~Jackiw and S.~Y.~Pi,
  ``Creation and evolution of magnetic helicity,''
  Phys.\ Rev.\  D {\bf 61}, 105015 (2000)
  [arXiv:hep-th/9911072].
  %%CITATION = PHRVA,D61,105015;%%
  
    
    \bibitem{adler}
  S.~L.~Adler,
  ``Massless Electrodynamics On The Five-Dimensional Unit Hypersphere: An
  Amplitude - Integral Formulation,''
  Phys.\ Rev.\  D {\bf 8}, 2400 (1973)
  [Erratum-ibid.\  D {\bf 15}, 1803 (1977)].
  %%CITATION = PHRVA,D8,2400;%%
   
  \bibitem{jackiwrebbi}
  R.~Jackiw and C.~Rebbi,
``Conformal properties of a Yang-Mills pseudoparticle,''
  Phys.\ Rev.\  D {\bf 14}, 517 (1976),
  %%CITATION = PHRVA,D14,517;%%
  ``Spinor analysis of Yang-Mills theory,''
  Phys.\ Rev.\  D {\bf 16}, 1052 (1977).
  %%CITATION = PHRVA,D16,1052;%%
  
  \bibitem{bogomolny}
    E.~B.~Bogomolny and Yu.~A.~Kubyshin,
  ``Asymptotical Estimates For Graphs With A Fixed Number Of Fermionic Loops In
  Quantum Electrodynamics. The Extremal Configurations With The Symmetry Group
  O(2) X O(3),''
  Sov.\ J.\ Nucl.\ Phys.\  {\bf 35}, 114 (1982)
  [Yad.\ Fiz.\  {\bf 35}, 202 (1982)].
  %%CITATION = YAFIA,35,202;%%
  
  \bibitem{jarvis}
  H.~S.~Booth, G.~Legg and P.~D.~Jarvis,
  ``Algebraic solution for the vector potential in the Dirac equation,''
  J.\ Phys.\ A  {\bf 34}, 5667 (2001)
  [arXiv:hep-th/0104216].
  %%CITATION = JPAGB,A34,5667;%%
  
  \bibitem{thooft}
  G.~'t Hooft,
  ``Computation of the quantum effects due to a four-dimensional
  pseudoparticle,''
  Phys.\ Rev.\  D {\bf 14}, 3432 (1976)
  [Erratum-ibid.\  D {\bf 18}, 2199 (1978)].
  %%CITATION = PHRVA,D14,3432;%%
    
  \bibitem{pais}
  A.~Pais,
  ``Spherical spinors in a  Euclidean 4-space'', 
  Proc.\ Nat.\ Acad.\ Sci.\ {\bf 40}, 835 (1954).
  





  \end{thebibliography}

\end{document}